\documentclass[sigconf, nonacm]{acmart}

\usepackage{balance}  % for  \balance command ON LAST PAGE  (only there!)
\usepackage{multirow}
\usepackage{here}
\usepackage{hhline}
\usepackage{url}
\usepackage{times}
\usepackage{listings}
\usepackage{float}
\usepackage{multicol}
\usepackage{algorithm}
\usepackage{algorithmicx}
\usepackage{algpseudocode}
\usepackage{here}

\if{}
\newcommand\vldbdoi{XX.XX/XXX.XX}
\newcommand\vldbpages{XXX-XXX}
\newcommand\vldbvolume{15}
\newcommand\vldbissue{1}
\newcommand\vldbyear{2021}
\newcommand\vldbauthors{\authors}
\newcommand\vldbtitle{\shorttitle} 
\newcommand\vldbavailabilityurl{http://vldb.org/pvldb/format_vol14.html}
\newcommand\vldbpagestyle{plain} 
\fi{}

\begin{document}
\title{ 
Serializable HTAP with Abort-/Wait-free Snapshot Read  
 }

%%
%% The "author" command and its associated commands are used to define the authors and their affiliations.
\author{Takamitsu Shioi}%}
\affiliation{%
  \institution{Department of Computer Science, School of Computing in \\ Tokyo Institute of Technology}
  \streetaddress{2-12-1 Oh-Okayama, Meguro-ku}
  %\city{Tokyo}
  %\state{Meguro}
  %\country{Japan}
  \postcode{43017-6221}
}
\email{shioi@de.cs.titech.ac.jp}

\author{Takashi Kambayashi}%}
\orcid{0000-0002-1825-0097}
\affiliation{%
  \institution{Nautilus Technologies, Inc.}
}
\email{kambayashi@nautilus-technologies.com}

\author{Suguru Arakawa}%}
\orcid{0000-0001-5109-3700}
\affiliation{%
  \institution{Nautilus Technologies, Inc.}
}
\email{arakawa@nautilus-technologies.com}

\author{Ryoji Kurosawa}%}
\affiliation{%
    \institution{Nautilus Technologies, Inc.}
}
\email{kurosawa@nautilus-technologies.com}

\author{Satoshi Hikida}%}
\affiliation{%
 \institution{%School of Computing in \\
   Tokyo Institute of Technology}
}
\email{hikida@de.cs.titech.ac.jp}

\author{Haruo Yokota}%}
\affiliation{%
  \institution{Tokyo Institute of Technology}
}
\email{yokota@de.cs.titech.ac.jp}

\begin{abstract}

Concurrency Control (CC) ensuring consistency of updated data is an essential element of   OLTP  systems. 
Recently, hybrid transactional/analytical processing (HTAP) systems developed for executing OLTP and OLAP  have attracted much attention.
The OLAP side CC domain has been isolated from OLTP's CC and in many cases has been achieved by  snapshot isolation (SI) 
  to establish HTAP systems. 
Although more high isolation level is ideal, 
 considering OLAP read-only transactions in the context of OLTP scheduling achieving serializability forces aborts/waits and would be a potential performance problem.
Furthermore, executing OLAP without affecting OLTP as much as possible is needed for HTAP systems.

The aim of this study was serializability without additional aborts/waits.  
We propose   % a model 
 read safe snapshot (RSS)  using multiversion CC (MVCC) theory  
  and introduce the RSS construction algorithm utilizing serializable snapshot isolation (SSI).   
For serializability of HTAP systems,  
 our model makes use of multiversion and allows more schedules with read operations whose corresponding write operations do not participate in the dependency cycles.
Furthermore,  we implemented the algorithm practically in an open-source database system that offers SSI. 
Our algorithm was integrated into two types of architecture as HTAP systems called as unified or decoupled storage (single-/multinode) architecture. 
We evaluate the performance and abort rate of the single-node architecture where SSI is applicable. 
The multi-node architecture was investigated for examining the performance overhead applying our algorithm. 
%by applying our algorithm and examining the performance overhead. 
%

\end{abstract}

\maketitle

\if{}
\pagestyle{\vldbpagestyle}
\begingroup\small\noindent\raggedright\textbf{PVLDB Reference Format:}\\
\vldbauthors. \vldbtitle. PVLDB, \vldbvolume(\vldbissue): \vldbpages, \vldbyear.\\
\href{https://doi.org/\vldbdoi}{doi:\vldbdoi}
\endgroup
\begingroup
\renewcommand\thefootnote{}\footnote{\noindent
This work is licensed under the Creative Commons BY-NC-ND 4.0 International License. Visit \url{https://creativecommons.org/licenses/by-nc-nd/4.0/} to view a copy of this license. For any use beyond those covered by this license, obtain permission by emailing \href{mailto:info@vldb.org}{info@vldb.org}. Copyright is held by the owner/author(s). Publication rights licensed to the VLDB Endowment. \\
\raggedright Proceedings of the VLDB Endowment, Vol. \vldbvolume, No. \vldbissue\ %
ISSN 2150-8097. \\
\href{https://doi.org/\vldbdoi}{doi:\vldbdoi} \\
}\addtocounter{footnote}{-1}\endgroup

\ifdefempty{\vldbavailabilityurl}{}{
\vspace{.3cm}
\begingroup\small\noindent\raggedright\textbf{PVLDB Artifact Availability:}\\
% The source code, data, and/or other artifacts have been made available at \url{\vldbavailabilityurl}.
\endgroup
}
\fi{}

\section{Introduction}

%{\bf Motivation.}
%
 Most DBMSs have been traditionally divided into write-optimized or read-optimized systems, 
 and have focused on utilizing  %themselves
  the system architecture for each workload features. 
OLTP- or OLAP-style applications are categorized as distinct characteristics, where  short period transactions update small data correctly in OLTP and long period ad hoc queries aggregate large data in OLAP. 
% low latency 
%
Accommodating OLTP and OLAP components on one system is highly challenging for a long time  
 because continuous OLTP update propagation and fast aggregation containing the updated data are performance tradeoffs that resulted from the difference of 
 % distinct 
  workload properties and appropriate  architectures.    
Moreover, merging OLAP read-only transactions into OLTP transactions would cause an anomaly  analysis result that contains inconsistent data \cite{Fekete2004}.    
In recent years,  
 hybrid transactional/analytical processing (HTAP) systems have been developed to support  both  OLTP and OLAP workloads  \cite{Raza2020, Makreshanski2017, appuswamy2017case, DBLP:journals/debu/LeeMMFSPKG13, HyPerFork2011kemper, Neumann2015}. 
There is considerable interest in such systems for use in modern database applications such as real-time data analysis \cite{chattopadhyay2019procella, Oracle2015Real-time, Plattner2009}. 
In \cite{Unterbrunner2009Predictable}, a concrete use case of the airline industry was shown.    
 Moreover, 
 HTAP systems are categorized into either unified (single-node) or decoupled (multinode) storage architectures as the physical composition \cite{Raza2020}.  
Multinode architecture HTAP systems try to communicate continuously  the physical partitioned systems for solutions  against batch propagation from daytime OLTP systems to OLAP data warehouses  \cite{HTAPSurvey2017ozcan}.  
Recent studies have concentrated on improving data freshness for analyzing the most recently changed data by connecting to a write-optimized system. 
 %

%{\bf Research problem.}
%
Although  performance improvement of HTAP systems has major issues,  concurrency control (CC) guaranteeing the consistency of the updated data also has been a vital aspect. 
HyPer-Fork \cite{HyPerFork2011kemper, HyPerFork2011Henrilk} has achieved  serializability based on   snapshot isolation (SI) \cite{2007Berenson} by  serial execution of OLTP transactions and snapshots  obtained  from the gap where concurrent update transactions are absent.  
To design serializability on both OLTP and OLAP engines,  however,  intentionally creating the gap on   the OLTP side would limit the concurrency performance of  OLTP write transactions.     
Thus, HyPer-MVOCC \cite{Neumann2015} aborts OLTP write transactions only when they concurrently modify the read-set of OLAP read-only transactions to achieve serializability on OLAP. 
We consider the ideal HTAP system should process OLAP without affecting OLTP performance. 
Such performance problems would arise from enhancing not only data freshness but also the isolation level as CC.

%{\bf Research problem.}
%
To achieve OLAP's serializability with little impact on OLTP performance, 
aborting write transactions on the OLTP engine would be unrealistic.
%
% \cite{Raza2020, Makreshanski2017}. 
% caldera2 {Raza2020} ,  BatchDB \cite{Makreshanski2017}. 
For utilization of a replicated system nondedicated to HTAP, conflict-prevention with read-only optimization (CP-ROO) \cite{Jung2011, Bornea2011} 
%   RSSI (Replicated Serializable Snapshot Isolation) {Jung2011} {Bornea2011} 。
 allows all read-only transactions to commit and validate the detected  read-write conflicts (rw-conflicts)  \cite{Adya2000},   satisfying the conditions for executing aborts of write transactions ({\it writer-aborts}).      
In the case of processing HTAP workloads, HTAP systems would deteriorate the OLTP performance because of  many writer-aborts by OLAP analytical queries scanning a larger read-set of data. 
Furthermore, communication overheads collecting read-write dependecies (rw-dependencies) for writer-aborts affects OLTP and OLAP performance on multinode architecture.  
Consequently, many single-/multinode HTAP systems have executed OLAP read-only transactions under SI for preserving OLTP performance instead of serializability using writer-abort or two-phase locking (2PL).  
To achieve serializability as HTAP systems,  some techniques would be required.
For example, 
PostgreSQL provides {\it safe snapshots} \cite{PortsDan2012} by setting  READ ONLY DEFERRABLE flags in transactions.    
 The {\it read-only deferrable transaction } forces 
  OLAP read-only transactions to wait ({\it reader-wait}) for the arrival of serializable snapshots that are  taken intentionally  by writer-abort  or  
%unintentionally (Safe Snapshots \cite{PortsDan2012} )    
% 
 delaying the concurrent write transaction starts ({\it writer-wait})   
% Silo {Silo2013}, Safe Snapshot {PortsDan2012}
  to obtain successfully serializable snapshots (such as the absence of concurrent writer on HyPer-Fork or Snapshot Epoch \cite{Silo2013} ).
However, 
 the wait time of read-only or write transactions depends on OLTP workloads and degrades system performance.   
 %
%

%{\bf Contributions.}
%
This study focuses on accomplishing serializability for HTAP systems.
Moreover,  our approach prevents read-only transactions from waiting  {\it while  executing  transactions in real time on OLTP}.   
We introduce {\bf read safe snapshot (RSS)} as the theoretical framework on top of  multiversion CC (MVCC) \cite{Adya2000, DBLP:books/aw/BernsteinHG87} to achieve serializability read-only transaction joins concurrently. 
% 
%{\bf Contributions.}
RSS can execute OLAP queries mapping previous versions without additional aborts/waits  of read-only transaction participation.  
Furthermore, we propose an RSS construction algorithm and utilize SSI \cite{2005fekete} to reduce RSS construction overhead in tracking transactional dependencies.    
Our proposed method  was implemented in PostgreSQL as a system offering SSI and the log-shipping replication technique for constructing an OLTP component and the read-only replica. 
% %
We investigated the performance of our implementation under CH-BenCHmark \cite{Cole2011, Difallah2013OLTPBench}  as % OLTP and OLAP
 workloads for evaluation.  
%
%In summary, we make the following contributions: 
%In particular, the main contributions of this paper are: 
The contributions of this study are:

%\vspace{-1.5\baselineskip} 
\begin{itemize}
%
% of MVCSR for read-only transaction participation.
\item Arranging a theoretical framework (RSS) 
%on conflict-serializability class \cite{Adya2000, DBLP:books/aw/BernsteinHG87}
to guarantee that a multiversion scheduler can accept  read operations  targeting shortly-previous-versions  when read-only transactions participated. 
The scheduler acceptability shows that additional aborts or waits caused by read-only transaction participation are not required to achieve serializability. 
\item Based on that, formalizing an algorithm reducing the cost of tracking transactional dependencies  to collect only rw-dependency information   
 in the case an OLTP engine applies SSI  in HTAP systems.   

\item  % Constructing 
Integrating  
 the algorithm in two types of HTAP architectures categorized as unified/decoupled (single-/multinode) storage  systems. 
It shows our theoretical framework can apply to a wide range of HTAP architectures.
Our implementation used PostgreSQL and the log-shipping replication function for the multinode  architecture. 

\begin{itemize}
\item  
 No matter what RSS applies multinode architecture,      
 RSS can achieve serializable (abort-/wait-free) in read-only replica by  constructing RSS beforehand through log-shipping of collected transactional dependency on the OLTP engine. 
\end{itemize}

\item Enhancing PostgreSQL's OLTP throughput up to 20\%    
 comparing SSI + SafeSnapshots  
 in unified architecture on the situation that OLTP clients increased to 48.
 On the other hand,  RSS OLAP performance does not decrease  in comparison with  SSI,  SSI+SafeSnapshots even if the number of OLAP clients increases. 
 In addition, compared with  SSI+SI 
 % unsupported serializability
  multinode architecture that   
  OLAP transactions executed under SI on PostgreSQL,  
 the demonstrations of 
 multinode architecture restrain OLTP throughput about 10\% in comparison with  SSI+SI. 
  % when the difference is the largest. 
 %
 RSS OLAP performance on multinode architecture  about 10\% lower than SSI+SI.   %SSI+SafeSnapshots and SSI+SI  even if OLAP clients increases.    

\end{itemize}

{\bf Outline.}
The rest of this paper is organized as follows.
Section 2 describes related works achieving SI as CC on  % a HTAP system 
 an OLAP engine.   
In Section 3, we introduce the notations and an organization of read-only transaction anomaly on HTAP.  
In Section 4, we consider the basic model and algorithm for ensuring HTAP serializability under the assumption that OLTP transaction history guarantees serializable when read-only transactions join in  real time.      
Section 5 presents the architecture implementing the presented algorithm based on PostgreSQL. 
Section 6 investigates the performance overheads of the prototype implementation to achieve serializability.  
In Section 7, we conclude this paper with future work.     
% 
%All proofs of theorems are given in the paper's Appendix.

\section{BACKGROUND and RELATED WORK} \label{sec:2}

As mentioned in the previous section, the aim of this study was the serializability of  HTAP systems without deteriorating performance and data freshness as much as possible. 
HTAP systems established the compatibility of the performance and data freshness, whereas  the OLAP engine compromises serializability. 
In the systems adopting SI,  a write transaction can update while a potentially long-running read transaction of HTAP workload proceeds concurrently. 
Unfortunately, SI does not always guarantee serializable behavior; it leads to inconsistent data, for example,  read-only transaction anomaly  \cite{Fekete2004}. 
%Feket ROA {Fekete2004}。
A generic HTAP system that is realized by SI  has isolated the CC domain from OLAP: the OLTP engine should be able to achieve serializability with traditional CC studies; however, the OLAP engine cannot achieve it.  

\subsection{SI-based HTAP systems} 

SAP HANA \cite{DBLP:journals/debu/LeeMMFSPKG13}  produced a consistent view manager that supported  SI  in the distributed in-memory database.  
%SAP HANA architecture1 {DBLP:journals/debu/LeeMMFSPKG13} 。
The transactions accessing the consistent view manager reads the most recent committed versions at  the time when the transaction began for ensuring the consistent view of SAP HANA, 
 and the two-phase commit (2PC) protocol ensures atomicity of distributed multi-node update transactions for SI.  
The write transactions would assure serializability if the 2PC protocol checks additionally quasi rw-conflicts using lock and aborts the conflicting transaction.
Although the short-running read-only transactions can reuse a cached consistent view to avoid performance bottleneck,  reading only old data cannot ensure serializability. 
Long-running read-only transactions also can ensure SI because of  the two-phase commit protocol: the read-only transactions are SI derived from a consistent view.

BatchDB \cite{Makreshanski2017} is primarily based on replication explored to achieve performance isolation for OLTP and OLAP workloads and  implied  SI  to realize an HTAP system.   
Such architecture could  isolate OLTP and OLAP workloads across hardware boundaries, as a result  from  each replica could apply systems dedicated for each workload;  the OLTP replica  implemented by Hekaton \cite{10.1145/2463676.2463710}. 
% SQLserver  Hekaton origin, OCC   \cite{10.14778/2095686.2095689, 10.1145/2463676.2463710}.
Hekaton MVOCC guarantees that the transactions  can achieve serializability by aborts with re-read validation and  write-write conflicts (ww-conflicts) detection \cite{Larson2011}.  
Considering performance overheads of serializability, the OLTP component in BatchDB  was  not used for  Hekaton's serializable isolation.  
In contrast, the OLAP component attained unlocking SI in the replicated system and maintained single version data individually  copied from  one batch.    
To keep consistent views, the OLAP replica  executed transactions  nonconcurrently while  executing  the batch propagation.  

\subsection{SI-based serializability}

HyPer \cite{HyPerFork2011kemper, HyPerFork2011Henrilk, Neumann2015}  achieved SI-based   serializability firmly considering the OLAP component. 
HyPer-Fork \cite{HyPerFork2011kemper, HyPerFork2011Henrilk}  processed OLTP transactions serially and OLAP queries on a consistent snapshot that resulted from Copy-on-Write between two serial transactions.  
HyPer-MVOCC \cite{ Neumann2015}  avoided write-write conflicts by lock and determined the visible version of a record by comparing the transaction's start timestamp to the commit timestamp of the version delta in the newest-to-oldest direction to achieve SI.  
Moreover, HyPer-MVOCC executed  writer-aborts to prevent  rw-antidependencies \cite{2005fekete}   from occuring to achieve serializability under SI.   
% SSI Feket {2005fekete}
HTAP systems largely  utilize snapshot mechanisms for CC and can take advantage of serializability like HyPer. 
Although some HTAP systems could exploit serializability, degrading the OLTP performance would be the cause of the supplementary OLAP transactions; in particular,  serial execution (writer-wait) or writer-abort may be expensive.

There are alternative ways of using snapshots to enable serializable isolation.     
SSI \cite{2005fekete} is known as a theory offering serializable isolation under SI.    
the nonserializable state of SI transactions contains two successive concurrent rw-antidependencies.
If a system that uses SI aborts  one of the transactions consisting of two successive rw-antidependencies, the system can achieve serializability. 
However, applying SSI to HTAP systems leads to  OLAP performance deteriorating by the reader-abort.   
Because OLAP's read-only transactions tend to be huge read sets and have long lifetimes based on  the characteristics of scan-mostly and  long-running  analytical queries,   
detections of the concurrent rw-dependencies would be time-consuming.  
In particular,  recent HTAP systems using replication need to send transaction's read sets from the read replica dedicated for OLAP workload to the OLTP engine, and concurrent write transactions with the long-running read transactions must  keep track of (over-)writes.
If two short-running update transactions are committed as the first rw-dependency before  the read-only transaction creates the second rw-dependency,  reader-abort of the second rw-dependency would be executed.   
In addition, if  s long-running read-only transaction  is retried, reader-abort would   easily occur  by   validation.    
HyPer-MVOCC  \cite{ Neumann2015} and CP-ROO \cite{Jung2011, Bornea2011} would yield a  greater OLAP performance than a system that applied SSI because the validation chooses writer-abort  instead of reader-abort when it detects the rw-conflicts.  
Therefore, applying SSI for the OLAP component is unsuitable for processing  OLTP and OLAP on multinode architecture.

In  the read-only replica of log-shipping replication,  Ports and Grittner \cite{PortsDan2012}  proposed   taking snapshots (safe snapshots) for serializable isolation based on SI.       
% 
%In order that
 Because 
  read-only transactions  under SSI can read a snapshot without the need to track rw-conflicts, 
  safe snapshots were exploited in a system that  offers SSI. 
 \cite{PortsDan2012}  illustrated the idea of taking reliably safe snapshots at read-only replica of log-shipping replication by writer-abort/-wait on an OLTP engine and reader-wait on the read-only replica.    
That is, such  transactions called  deferrable transactions in the system  must abort concurrent write transactions, or must prevent  new transactions from starting while waiting for concurrent write transactions to finish; 
moreover, 
read-only transactions on the replica must wait for safe snapshots to arrive.     

\section{Preliminary: problem statement} \label{sec:PM}  % research question, problem statement  

This section introduces the formalization that is necessary for our study, and illustrates a concept that ensures global (conflict-) serializability \cite{Schenkel2000SMVSG, Wachter1992LSRGSR} under the theory of classical MVCC with the additional restriction on OLAP read-only transactions:  
the read-only transactions do not  know the conflicts with Active transactions until the write  transactions are committed.  
In particular,  waiting for  the confirmed information would  deteriorate the performance for real-time analysis on HTAP systems;    
we define a problem structure caused from read-only transaction participation in Section  \ref{sec:roa}.  
In Section \ref{sec:rss},  by considering the problem structure, we suggest a method identifying the committed transaction region that read-only transactions can read the  previous-versions guaranteed  to achieve serializability.  

\subsection{HTAP scheduler as a global serializability} 
% Papa  \cite{1986papadimitou} 
%  Bernstein \cite{DBLP:books/aw/BernsteinHG87}  
% Adya et al. \cite{Adya2000}  

%
 Adya et al. \cite{Adya2000, adya99:_weak_consis} have introduced Isolation Level PL-3 ensuring serializability within the class of  conflict-serializability (Bernstein et al. \cite{DBLP:books/aw/BernsteinHG87} ).      
In this paper, we adopt the multiversion history formalization in \cite{Adya2000} and assume transaction schedules/histories are always given with a version-order of writes. % 
We call the class of multiversion schedules satisfying serializability (PL-3 \cite{Adya2000})  as version-ordered conflict-serializability (VOCSR).  
The history of an OLTP engine is assumed to be serializable in the meaning of VOCSR, 
and OLTP engine assumes that read operations can read only committed versions.
HTAP scheduler processes operations in transactions as a dynamic scheduler that implies isolating each OLTP and OLAP protocol. 
HTAP history (or complete schedule) is a sequence that consists of the union of the operations from given transactions on each OLTP and OLAP.   
The prefix of operations that have been executed by now in the ongoing schedule is denoted the     current prefix.  
We design HTAP history acceptable to HTAP schedulers to be spontaneously global serializability, what we call serializable HTAP, ensuring the following conditions;

\begin{itemize}
\item OLTP does not perceive the processing contents of OLAP: 
 when appending an OLTP operation, OLTP protocols  know only the OLTP operations in the current prefix  of a history.

\item OLAP can perceive the processing contents of OLTP (through log-shipping or other processes) : 
  when appending an OLAP operation, OLAP protocols  know the committed transactions on the OLTP component as well as the OLAP component operations in the current prefix of the history, including version-order defined by the ww-conflicts  in VOCSR.  

\end{itemize}

\noindent
Note that the VOCSR can be considered to contain a history SSI generates as serializability class, as the version-order is induced by the order of the commit operations of the transactions that wrote the versions (SI version order  \cite{Schenkel2000Weikum} ).  

 \subsection{Notation}

Let $ \mathbb{T} = \{ T_1, \dots , T_n \} $ be a set of transactions.
A transaction $T$ is a (totally ordered) sequence  of read and write operations on data items.  
A write operation on data item $X$ by transaction $T_a$ is denoted by $W_a(X_a)$; if it is useful to indicate the value $v$ being written into $X_a$, we use the notation, $W_a(X_a, v)$ \cite{adya99:_weak_consis}.      
When a transaction $T_a$ reads a version of $X$ that was created by $T_b$, we denote this as $R_a(X_b)$. 
If it is useful to indicate the value $v$ being read, we use the notation $R_a(X_b, v)$.     
Here,  about a history $h$ and such operations $o_i, o_j \in h$, we denote $o_i \prec_h o_j$, if $o_i$ preceded $o_j$ in $h$. 
For a transaction $T \in \mathbb{T}$, we write $Begin(T)$ as an operation that appears in most preceded operations of $T$  on  a  history. 
Similarly,  we write $End(T)$ as the most successor in $T$  on the history $h$;   
$Begin(T) \prec_h End(T)$  always holds.

%SI-V, SI-W
Schenkel and Weikum \cite{Schenkel2000Weikum, 2001weikum} have defined SI as  a multiversion scheduler that holds two conditions. 
For later explanation, we interpret and describe the property of SI.   
The first condition denoted as SI-V (SI version function) is a read protocol that maps each read  operation $R_a(X_b)$ to the most recent committed write operation $W_b(X_b)$ that begins at 
  %as of the time of the begin of
 $T_a$. 
The second condition defined as SI-W (disjoint writesets) is  a commit or write protocol (called as First Committer/Updater Wins Rule) where the writesets of two concurrent transactions are disjoint.

Adya et al. \cite{Adya2000} have defined the direct serialization graph (DSG) arising from any given  multiversion history $h$. 
Each node in DSG(h)  corresponds to a committed transaction and directed edges correspond to   conflicts (write-write, write-read, read-write)  under  information about committed transactions (a committed projection of  the current prefix in $h$).  
Such a transaction order (partially ordered) graph guarantees serializable from the given history if it is acyclic.   
For  transactions $T_a, T_b \in \mathbb{T}$,    we write $T_a \to T_b$ (dependency edge)  if there exists any of the direct conflicts (write-write, write-read, read-write) from $T_a$ to $T_b$.   
Let us write $T_a \to^* T_b$  for the reflexive and transitive dependency where $T_b$ is reachable from $T_a$; note that  $T \to^* T$ always holds for any transaction $T$, but it is not considered as a cycle. 
% unreachable dependency 
We say unreachable if no directly or transitive dependency edge from $T_a$ to $T_b$ holds then we denote $T_a \not\to^* T_b$. 
%We denote that directly or transitive dependency 
%$T_a \not\to^* T_b$ where $T_b$ is unreachable from $T_a$ as directly or transitive dependency. 

\subsection{  The missing piece: read only anomaly on OLAP
 %Problem structure: Read-only anomaly
} \label{sec:roa}

Here, we organize a problem resulting from read-only transaction participation.
To achieve serializability as HTAP in our precondition, we aim to define and remove a set of anomalies  that include read-only transaction anomalies \cite{Fekete2004} . 
% 
%In particular, 
Even if the OLTP side of an HTAP scheduler is serializable, an anomaly situation  would occur on the  OLAP side.    
We illustrate such problems on HTAP systems in this section.

\begin{definition} Read-Only Anomaly 

\noindent
We say A read-only anomaly occurs if a history $H$ over transactions $ \mathbb{T}$ contains nonserializable structure $\mathbb{S}  \subseteq   \mathbb{T}$, where a set of transactions $\mathbb{S} = \{ S_1,\ S_2,\ S_3,\ \cdots ,\ S_n \}$  holds the following conditions;

%
%H() ha 
\begin{itemize}

\item History $H$ restricted over transactions $\{S_1,\ \cdots \ ,\ S_{n-1} \}$ is a serializable history in  VOCSR: 

 $ H( S_1,\    \cdots \ ,\ S_{n-1} )  \subseteq  VOCSR$,        

\item $ S_{n}  $ is a read-only transaction, and     

\item  $ \mathbb{S} $ has the cycle of dependency edges $P$;     
     
$ P= \{ S_1  \to^*  S_2,\  S_2  \to^*  S_3,\  \cdots, \   S_{n-2}  \to^*  S_{n-1}, \  S_{n-1}  \to S_n, \  S_n  \to  S_1 \}$. Note that the last two edges $S_{n-1}  \to S_n$ and $ \  S_n  \to  S_1 $ are direct dependency. 

\end{itemize}

\end{definition}

In history $h_s$ under SI, we describe  the example referred to as read-only anomaly  \cite{Fekete2004} :

\begin{flalign}
 h_s :   R_2(X_0, 0) \  R_2(Y_0, 0)  \  R_1(Y_0, 0) \  W_1(Y_1, 20)  \nonumber   \\
   R_3(X_0, 0)  \  R_3(Y_1, 20)   \  W_2(X_2, -11)    \nonumber  
\end{flalign}

\noindent  
Although the history over $T_1$ and $T_2$ is serializable under SI,  nonserializable history $H$ is caused by participating read-only transaction $T_3$: 
$h_s$ is a read-only anomaly because of  rw-dependency edges of $T_2 \to T_1$ and $T_3 \to T_2$ and  transaction $T_3$ is read-only transaction arising write-read dependency (wr-dependency)  edge $T_1 \to T_3$.   
If $h_s$ arises under SSI, a transaction would be aborted because of dangerous structure in the rw-dependency edges of $T_2 \to T_1$ and $T_3 \to T_2$.
Transactions executed under SSI  do not arise 
 %occur under
  read-only anomalies.

In the case of an HTAP scheduler, let $\mathbb{T}$ be accepted under SSI
 and $T_3$ be an OLAP read-only transaction. 
 % read-only transaction of OLAP.   
%
In  the OLTP side, the protocols interpreted the sequence of operations in $T_1$ and $T_2$,  $\{ R_2(X_0) \  R_2(Y_0)  \\ \  R_1(Y_0)   W_1(Y_1) \  W_2(X_2) \}$ are accepted under SSI, and $T_1$ and $T_2$ can commit without considering the OLAP side.   
%
%On the other hand, 
 In contrast, 
 if the read-only transaction $T_3$ processes are on the OLAP side,  where the current prefix is between $End(T_1)$ ($= W_1(Y_1)$ ) and $End(T_2)$ ($= W_2(X_2)$ ), then the read protocol SI-V    reads the version $Y_1$ and $X_0$.
Hence,  $T_3$ would be reader-abort of the second rw-dependency in dangerous structures because  the OLTP side would not writer-abort the already committed transactions $T_1$ and $T_2$ for performance by OLAP  circumstances.    
In fact, validating reader-abort on SSI must wait for  the information of $End(T_2)$ through committed log, and  retrying the read-only transaction while processing successively write transactions would abort again: reader-abort and retry are inefficient for HTAP systems.

To achieve serializable isolation without aborts, there are two types of approaches.   
In the $h_s$, 
 reading the {\it most recent} committed version $Y_1$ on the current prefix of OLAP side resulted in  a read-only anomaly. 
If the read protocol of $T_3$ chooses the {\it previous} version $Y_0$,   the scheduler cannot have led to the read-only anomaly nor aborted a transaction for serializability.    
Otherwise,  even if $T_3$ read the {\it successor} version of $X_2$ instead of $X_0$, 
 $T_3$ needed to (read-) wait the commit of $T_2$ not knowing when it will end.  
Moreover, concurrent transactions is not aware of whether or not the successor version $X_2$ is written,   and  the concurrent rw-dependency  can arise after such the validation similar to $End(T_3)$ ($=R3(Y1)$ )   $ \prec_{h_a} W_2(X_2)$.      
Therefore, we focus on constructing the model reading {\it previous} versions ensuring serializability. 
%without an abort. 
% 
%
We consider that the key of MVCC is read protocols that can be free from restrictions on read versions (not necessarily the most recent one).

\section{Read Safe Snapshot (RSS)} \label{sec:rss}

Ideally DBMS would have little effect on performance to achieve serializable isolation.   
As we described in Section. \ref{sec:2}, 
previous studies have provided SI to read-only transactions regardless of whether it assures serializability.
Traditionally,  guaranteeing serializability is more desirable than tolerating read-only anomalies under SI,   but the methods were far from optimal as utilizations for read-only transaction participation,    
%\cite{2005fekete, HyPerFork2011kemper, Neumann2015, Jung2011, PortsDan2012} ,    
 because  potential performance overhead comes from aborts or waits to take and send the snapshot written nonconcurrently.    
% \cite{Neumann2015, Jung2011} 
% \cite{HyPerFork2011kemper, PortsDan2012}  

% 
In this section, we improve a model for constructing a view (snapshot) assuring serializability whenever  read-only transactions join. 
The model uses the following concept;  

\begin{itemize}    
\item 
First, because 
%Firstly, on the grounds that 
 a read-only transaction perceives beforehand  that read-only anomalies do not occur by reading the prepared view, the read-only transaction does not need to do additional validations like finding an rw-conflict, causing the writer-/reader-abort.

\item 
Second, the read-only transaction does not perceive what and when concurrent transactions will  write or commit.       
Therefore, our model  identifies the boundary between the concurrent and finished transactions at any point (prefix) of the history,   
 and creates  in advance  the view from unreachable dependency edges. 
  In fact, if the read-only transactions always read the prepared view in transactional processing, they do not make OLTP/OLAP transactions wait.
  
\end{itemize}

\noindent 
The features obtained from our model are serializability, with no additional validations and the aborts/waits concurrently executing write transactions,       
 allowing read-only transactions to participate in real-time analysis like the HTAP application. 

\subsection{Theoretical framework} \label{model}

\begin{figure}[h]
 \begin{center}
 %\vspace*{2cm}
  \includegraphics[bb=81 119 800 473, clip,width=1.0\linewidth]{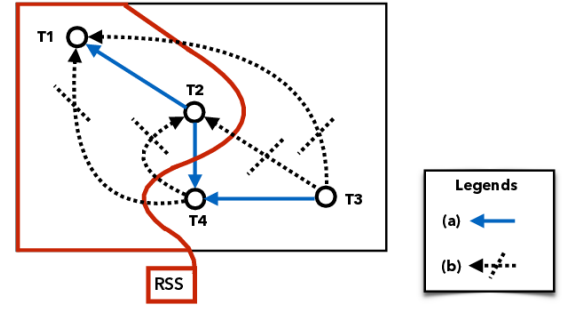}  % bb=184 128 795 468
  \caption{
  Conceptual diagram of RSS.    
  (a) represents dependency edge. 
 (b) represents transactions are unreachable in the direction of the arrow because  conflicts do not occur on the history prefix.
  From a read-only transaction point of view, $T_1$ and $T_2$ in RSS can be considered as a transaction region protected against dependency paths through $T_3$ or $T_4$.
    }
  \label{fig:RSS_concept}
 \end{center}
\end{figure}

In what follows,  we introduce read safe snapshot (RSS) as a formal definition,  
where we use the concept of conflicts on VOCSR derived from the model of Adya et al. \cite{Adya2000} as the prerequisite described in the previous section.   
To obtain a set of transactions contained in the serializability class, RSS is defined as an unreachable transaction region on the conflicts,  as shown in Figure \ref{fig:RSS_concept}.    
Moreover, we define a property with respect to the set of read-only transactions outside RSS, 
where the operations make read-targets the most successor versions in RSS.       
Finally, we claim that a multiversion scheduler achieving serializability  can correctly accept operations of the read-only transactions in Theorem \ref{theorem:ptr}. 
% on theorem \ref{theorem:ptr}.   

% Figure: RSS concept 

% RSS % Protected  Transaction Retion (PTR) \label{PTR} 
\begin{definition} (Read Safe Snapshot, RSS)  \label{RSS} 

\noindent 
Let $\mathbb{T}$ be a set of committed transactions, $\mathbb{P}$  be a set of transactions 
contained in a history $\subseteq$ VOCSR, 
and  $\mathbb{P} \subseteq \mathbb{T}$.  
We say that 
$\mathbb{P}$ is RSS, where $\mathbb{P}$  satisfies the following condition; 
%protected transaction region (PTR) 

\begin{itemize}

\item for arbitrary transactions $T_p \in \mathbb{P}$ and $T_q \notin \mathbb{P}$ (hereinafter $T_q \in \mathbb{T}$ for $T_q$), 
 $T_p$ is unreachable from $T_q$: $T_q \not\to^* T_p$.   

\end{itemize}

\end{definition}

% Protected Read only Transaction (PRoT) \label{RSS} 
\begin{definition} \label{PRoT}  (Protected read-only transactions regarding to $\mathbb{P}$, $R(\mathbb{P})$ )   

\noindent
Let  $\mathbb{T}$ be a set of committed transactions, $\mathbb{P}$ be RSS in $\mathbb{T}$,
and $T \in \mathbb{T}$, where $T$ 

\begin{itemize} 

\item is not contained in $\mathbb{P}$;    

\item has no write operations;  and      

\item has only read operations that read
 the versions created by write operations in most recent committed transactions in $\mathbb{P}$.      

\end{itemize}

\noindent 
 then, we say that $T$ is a protected read-only transaction (PRoT) regarding $\mathbb{P}$  and denote $R(\mathbb{P})$ about  this property.   

\end{definition}

We now introduce lemma \ref{lemma:1} to prove Theorem \ref{theorem:ptr}.  
The Lemma \ref{lemma:1} claims that the transactions holding $R(\mathbb{P})$ cannot have an incoming write-read conflict with regard to transactions excluded from the same set $\mathbb{P}$.      
In addition, Theorem \ref{theorem:ptr} claims that a scheduler regarded as a hybrid as VOCSR and another scheduler can be correctly acceptable in one history as globally serializable.

% lemma SSI1 
\begin{lemma}   \label{lemma:1}

 For any $T_r$ holding $R(\mathbb{P})$ and $T_q \notin \mathbb{P}$,  $T_r$ is unreachable  from $T_q$:  \  $T_q \not\to^* T_r$. 
    
\end{lemma}

\begin{proof}

We assume that $T_r$ is reachable from $T_q$, that is, there is the chain $T_q \to^* T'_q \to T_r$. 
Here, $T'_q \notin \mathbb{P}$ because otherwise, $T'_q \in\mathbb{P}$ is reachable from $T_q \notin \mathbb{P}$, contradicting the fact that $\mathbb{P}$ is RSS.
A conflict with destination $T_r$  only exists as a write-read conflict,  because $T_r$ do not include write operations  by definition \ref{PRoT}.   
However,   transactions  having  the write-read conflict with destination $T_r$ can  exist  only  on     transactions contained in $\mathbb{P}$.  
 Thus, $T_q \not\to^* T_r$ holds. 
    
\end{proof}

\begin{theorem} \label{theorem:ptr}
For a history $h$ that  applied committed projection, let $h$ contain $\mathbb{P}$ that compose RSS.     
  If VOCSR scheduler $s$ accepts   a history $h'$ that is the removal of the operations of $T_r$ holding $R(\mathbb{P})$ from $h$,  then $h$ is also accepted.  

\end{theorem}

 \begin{proof}
    In $h'$,  because of  Definition \ref{PRoT}  and Lemma \ref{lemma:1},      
    the lost conflicts from $h$ are limited to
    any of the following; 
  \begin{itemize}
  \item[(A)] 
  for a transaction $T_p \in \mathbb{P}$,  $T_p \to T_r$; or     
  \item[(B)] 
  for a transaction $T_q \notin \mathbb{P}$,  $T_r \to T_q$;   
  \end{itemize}
  
  \noindent 
 where we assume  that a dependency graph from given $h$ contains a cycle.    
The cycle contained a dependency edge of either (A) or (B) on the dependency graph from $h$,     because  a dependency graph from $h'$ does not contain cycles  by the definition \ref{RSS}.    
In this case of (A),  then $T_p \in \mathbb{P}$, and the cycle is $T_p \to T_r \to^* T_p$.  
Similarly, in the case of (B),  $T_q \notin \mathbb{P}$, and the cycle $T_r \to T_q \to^* T_r$ exists. 
In both cases,  about a $T'_p \in \mathbb{P}$ and  $T'_q \notin \mathbb{P}$, 
$T'_q \to T'_p$ or $T'_q \to T_r$  needs to appear  in the cycle, 
but that cannot hold by Definition \ref{RSS} and \ref{lemma:1}:             
  $h$  cannot  create  a cycle on a dependency graph.   
 Thus, if  $h'$ is accepted by $s$,  then  $h$ is also acceptable.   
\end{proof}

 \begin{corollary}
  The opposite is also true: if $h$ is accepted by $s$, then $h'$ is also acceptable.  
  % is accepted  by $s$, then  $h'$ is also acceptable.  
\end{corollary}

\begin{proof}
Because $h'$ is obtained by removing only some conflicts from $h$,  if a conflict graph computed from given $h$ has no cycle,  a graph from $h'$ is also acyclic. 
Thus,  $h'$ is accepted by $s$ in the case that $h$ is accepted by $s$.  
\end{proof}

%  summary 
As we described in this Section \ref{model}, our model shows that a history could not effect  scheduler acceptability  if  the scheduler removed transactions that hold $R(\mathbb{P})$.       
Because this manipulation is recursively applicable, it does not matter how many times such transactions are  eliminated from a history.  
In other words, as long as a read-only replica perpetually reads any one of some RSS is constructed at the backend of transactional processing,    
 the scheduler acceptability on the primary replica is not influenced.     
Finally,  if a primary replica gives attention to the scheduler acceptability under traditional MVCC on OLTP,
  achieving global serializability on HTAP (serializable HTAP) can be secured firmly. 
In fact, RSS would not  be capable of including the most recent committed versions, because OLTP that  keeps processing transactions and reachable dependencies against the versions would arise.
However, read-only transactions reading RSS could achieve  serializability even if the read-only transactions joined at all times while processing OLTP transactions.

\subsection{SSI-based RSS construction algorithm} \label{SHTAP}

In this section, we propose and inspect a construction method of RSS.   
The model we denoted in Section \ref{model} showed that read-only transactions could read versions ensuring serializability.       
However, a straightforward implementation requires information from each conflict for tracking a dependency graph.  
%would  require  informations of each conflicts for tracking a dependency graph.  
%      
Therefore,  naively implementing our model would cause potential performance problems for OLTP or OLAP  
 because it requires multiple complex steps; 
 for example,  extracting conflicts included in the nonconcurrent state on the OLTP component   increases  tracking paths in a dependency graph and  garbage collections  of conflict information to clean up stale RSS.
 % on the OLAP side. 
%   
We should consider how to implement our model without affecting OLTP performance by OLAP participation.

As we described in Section \ref{sec:2},  many previous  systems have been based on SI or implemented  other isolation levels.  
Because our model postulates serializability on the OLTP side, we focus on the SSI of a representative example for the serializability that consists of SI.   
For implementation, we propose an algorithm that optimizes our model in SSI on OLTP. 
The algorithm has the characteristics of extracting only concurrent rw-antidependency.  
In this section, we make a premise on the property of SSI (dangerous structure \cite{2005fekete} , SI-V and SI-W \cite{Schenkel2000Weikum} ) to develop a tractable algorithm.

Let us write the algorithm with a definition in advance.  
We define two types of a set of transactions and describe our algorithm as follows; 

\begin{definition} \label{txnState}  (Transaction state; $Done(p)$, $Clear(p)$ )  

  Let $h$ be a history under SSI, $p$ be its prefix, and $\mathbb{T}$ be a set of transactions contained in $h$.     
 We denote     
 
 \begin{itemize}
 \item
  $Done(p) = \{ T \in \mathbb{T} \mid End(T) \in p \}$; and 
 
 \item
   $Clear(p) = \{ T_a \in \mathbb{T} \mid \forall T_b \notin Done(p) $ such  that  $End(T_a) \prec_h Begin(T_b) \  \}$. 
  
 \end{itemize} 
  
\end{definition}

% arakawa & kurosawa algorithm   
\begin{algorithm}  
 \caption{RSS construction algorithm under SSI} 
  \label{algorithm1}

   \begin{itemize}

   \item[(1)] Contain entire $Clear(p)$ in RSS.     

   \item[(2)] Pick up any $T_c \in Clear(p)$.   

   \item[(3)] Pick up $T_u \notin Clear(p)$ and add to RSS if $T_u \to T_c$ exists.

   \item[(4)] Repeat Step (3) for all $T_u$

   \item[(5)] Repeat steps (2)--(4) for all $T_c$

  \end{itemize}

\end{algorithm}

\begin{figure}[h]
 \begin{center}
  \includegraphics[bb=60 80 802 525, clip, width=1.0\linewidth]{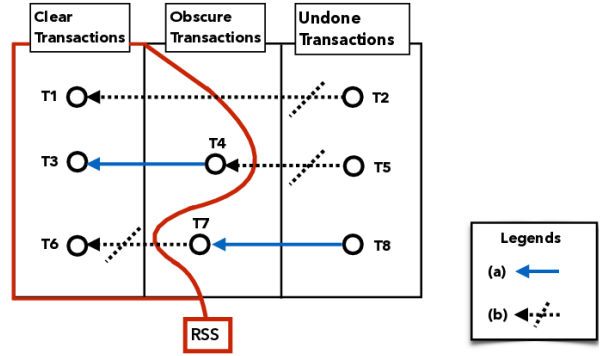} % bb=64 88 805 507
  \caption{
  Conceptual diagram of RSS construction algorithm under SSI.     
 (a) represents rw-dependency edges. 
 (b) represents transactions are unreachable in direction of the arrow as rw-conflicts;   
 $T_1$ and $T_2$ show that the nonconcurrent relation between Clear and Undone cannot give rise to  rw-dependencies.   
 $T_4$ and $T_5$ show that dangerous structure cannot arise:   
   if SSI scheduler had accepted $T_3 \leftarrow T_4$,     
   ended (committed) transaction $T_4$ cannot be aborted/reachable.   
 Similarly, 
  if $T_7 \leftarrow T_8$ holds, the SSI scheduler also should not have allowed   
    $T_6 \leftarrow T_7$, 
     thus, RSS cannot contain $T_7$ in spite of being committed.  
 $T_8$ is an uncommitted transaction; 
  as a result, 
  even if $T_7 \leftarrow T_8$ did not arise, 
   RSS could have contain $T_6$ 
   because  $T_6$  had been unreachable as Clear transactions.   
   }
  \label{fig:Alg_concept}
 \end{center}
\end{figure}

Algorithm \ref{algorithm1} utilizes our theoretical framework to construct RSS by SSI. 
In Figure \ref{fig:Alg_concept}, we show the relation between SSI properties and the set of transactions we defined.   
Here, Undone transactions (a complementary set of Done transactions, $T \in Done(p)^c$)    
  are not concurrent with Clear transactions $Clear(p)$ by the read protocol SI-V. 
% 
%
%Moreover, 
On the other hand, possible concurrent transactions with Clear transactions are contained in $Done(p)$. 
We call such transactions in $Done(p)$  Obscure transactions ($T \in Done(p)  \ \wedge  \  \notin Clear(p)$). 
For reaching from Undone transactions to Clear transactions, two dependencies through 
 %one of 
  Obscure transactions are essential.   
As shown in Figure \ref{fig:Alg_concept}, such essential features cannot arise as dangerous structures  under SSI. 
In summary, 
Undone transactions are unreachable to Clear and Done transactions having rw-dependencies between the Clear transactions.  
% by property of SSI. 
%
Thus, 
 on any prefix of an SSI history, 
 the unreachable transactions can join RSS, and read-only transactions can read the RSS  to achieve  serializability.

% Figure: RSS concept 

\subsection{Correctness of SSI-based algorithm} \label{correctness_alg}  %algorithm details 

% 
%If illustrations of our algorithm described in the previous Section \ref{SHTAP} are sufficient, 
%you can proceed to the next Section \ref{sec:implement} 
%because this section illustratee mostly theoretical details of Algorithm \ref{algorithm1}. 
% 
%The next section shows implementations with our algorithm on a system offering SSI.   

%
In what follows, we try to illustrate and inspect this algorithm.   
Here, we denote notations for later explanation. 
Let us organize self-evident properties derived from Definition \ref{txnState};   
 $Clear(p) \subseteq Done(p)$ holds:   
 for $\forall T_a \in Clear(p), T_b \notin Done(p)$, $End(T_a) \prec_h Begin(T_b)$ holds; 
 and 
for $\forall T_a \in Clear(p), T_b \notin Clear(p)$, $End(T_a) \prec_h End(T_b)$ holds.  
Furthermore,  we introduce definitions of dependency and dangerous structure in Fekete et al.   \cite{2005fekete}. 
For $T_a, T_b \in \mathbb{T}$, 
if operations of $T_a$, $T_b$ are  
concurrent  
 (  $Begin(T_a) \prec_h Begin(T_b) \prec_h End(T_a)$;          
  $Begin(T_a) \prec_h End(T_b) \prec_h End(T_a)$;  
  $Begin(T_b) \prec_h  Begin(T_a)$  $\prec_h End(T_b)$;   or   
   $Begin(T_b) \prec_h End(T_a) \prec_h End(T_b)$. )    
   and $T_a$, $T_b$ satisfy $T_a \to T_b$,   
  then we say that the conflict is vulnerable dependency. 
Note that  the conflict identified as a vulnerable dependency under SSI  only exists as concurrent rw-dependency.   
For $T_a, T_b, T_c \in \mathbb{T}$,
 if $T_a \to T_b \to T_c$ holds and both dependencies are vulnerable dependencies,   then this structure is dangerous structure. 
An SSI scheduler does not accept the schedule composing this structure.

We now denote some lemmas for verifying the appropriateness of an algorithm,   
 where  a history $h$ is committed projection and accepted by an SSI scheduler.
Lemma \ref{cb-order} claims that direct dependencies do not arise from Active transactions to  finished transactions  between nonconcurrent  transactions.    

\begin{lemma}  \label{cb-order} (SSI-1)  %(commit-begin order, SSI-1)

For a history $h$ and $T_a, T_b \in \mathbb{T}$ included in $h$, 
if $End(T_a) \prec_h Begin(T_b)$ holds, then $T_b \not\to T_a$ holds. 

\end{lemma}

\begin{proof}

Considering  each conflict,   
in the case of write-read conflict, that holds because of SI-V:
 the SI read protocol SI-V  can only map each read operation to the most recent committed write operations  when the self-transaction begins. 
 %  as of the time of the begin of self-transaction.  
%
Thus,  $T_b \to T_a$ cannot arise  write-read conflicts by $End(T_a) \prec_h Begin(T_b)$. 
In the case of write-write conflict, that holds because of  SI-W:
 SI-W was defined  so that the writesets of two concurrent transactions are disjoint, and SSI uses SI-W and obtains practically write version orders by the commit operation order;    
$T_b \to T_a$ cannot hold on write-write conflicts.  
In the case of read-write conflict, that holds because of  SI-V:
 if $T_b \to T_a$ had held,  then  $End(T_a) \prec_h Begin(T_b)$ must be contradiction by reading the most recent committed versions.  
Therefore,  for each conflict,  if $End(T_a) \prec_h Begin(T_b)$ holds, then the dependency is  unreachable from $T_b$  to $T_a$.  
 
\end{proof}

In Lemma \ref{conconflict}, we organize conflicts in the case that the  begin operation induced from processing any operations precedes (happens before) the end operation. 
 If the case is nonconcurrent,  each direct conflict cannot occur against the commit operation order by Lemma \ref{cb-order}.     
Otherwise,  read-write conflict only exists as the  reverse direction of transaction order between  concurrent transactions. 
Moreover, we describe Lemma \ref{ssi3} , which  claims the property of Lemma \ref{cb-order} related with  $Clear(p)$ of  Definition \ref{txnState}.  
The set of transactions in $Clear(p)$ is decided by the given prefix $p$ of the history, and  an  
 arbitrary prefix of the history can be used.

\begin{lemma}  \label{conconflict}  (SSI-2)  %(concurrent conflict, SSI-2) 

For a history $h$ and $T_a, T_b \in \mathbb{T}$ in $h$,  
let $Begin(T_a) \prec_h End(T_b)$.
If $T_b \to T_a$ holds, then the dependency is vulnerable dependency. 

\end{lemma}

\begin{proof}
If  $Begin(T_a) \prec_h End(T_a) \prec_h Begin(T_b) \prec_h End(T_b)$, 
 then $End(T_a) \prec_h Begin(T_b)$,  
  %where
     here 
   $T_b \to T_a$ does not hold by Lemma \ref{cb-order}.
By the definition of vulnerable dependency,  dependencies excepting this case exist only as vulnerable dependencies. 
Therefore, it holds.

 \end{proof}

 \begin{lemma}  \label{ssi3}  (SSI-3)

For a history $h$ and its prefix $p$, pick up an arbitrary $T_c \in Clear(p)$ and $T_u \notin Clear(p)$. 
If $T_u \to T_c$ holds, then it is a vulnerable dependency. 

 \end{lemma} 

\begin{proof}
   Because $End(T_c) \prec_h End(T_u)$  by $Clear(p)$ in definition \ref{txnState}, 
   $Begin(T_c) \prec_h End(T_u)$ holds.
   Thus, that holds from Lemma \ref{conconflict}. 

\end{proof}

We showed that the transactions contained in $Clear(p)$ are only reachable from transactions outside $Clear(p)$ through a vulnerable dependency in Lemma \ref{ssi3}.  
Although a
  %the one of 
  vulnerable dependency can exist in a dangerous structure (two successive vulnerable dependencies),  
an SSI scheduler cannot accept the second vulnerable dependency.
If  an SSI scheduler has accepted two successive dependencies that include a vulnerable dependency  as the first dependency, 
then the second dependency could not exist as a vulnerable dependency: 
write-write or write-read conflicts can exist as the second dependency. 
The ww-dependency and wr-dependency can arise on nonconcurrent states and  not 
 %create
 reverse the direction of transaction order.  
That is,  on that occasion,  transactions that created the outgoing second dependency edge are also contained in $Clear(p)$ by the definition of $Clear(p)$.

\begin{theorem}  \label{ssi4}  (SSI-4) 

For the SSI history $h$ and its prefix $p$, let $T_c \in Clear(p)$, $T_u \notin Clear(p)$, $T_v \in \mathbb{T}$. 
If $\ T_v \to T_u \to T_c$ holds,
 then $T_v \in Clear(p)$. 

\end{theorem} 

\begin{proof}
$T_u \to T_c$ is a vulnerable dependency by Lemma \ref{ssi3} and then a conflict exists. 
 By contraposition of Theorem \ref{cb-order},  $Begin(T_u) \prec_h End(T_c)$ holds.  
 Moreover, $T_v \to T_u$  leads to  $End(T_v) \prec_h Begin(T_u)$,  because the SSI scheduler does not accept a dangerous structure:
if $T_v \to T_u$ could exist, then an SSI scheduler must have accepted operations of nonconcurrent transactions.    
Thus,  $End(T_v) \prec_h End(T_c)$ holds and such $T_v$ can be contained in $Clear(p)$ only. 

\end{proof}

\noindent 
We now prove the validity of our algorithm.  
Assume that $\mathbb{P} \subseteq \mathbb{T}$ is  a subset detected by the Algorithm \ref{algorithm1}. 
 To say that $\mathbb{P}$ is RSS,  
for any $T_p \in \mathbb{P}$, $T_q \notin \mathbb{P}$, 
we need to show $T_q \not\to^* T_p$. 
To prove by contradiction, we assume $T_q \to^* T_p$. 
Then there must be at least one direct dependency in the chain from a transaction outside $\mathbb{P}$ to a transaction in $\mathbb{P}$.
By renaming the transaction if necessary, we assume $T_q \to T_p$, where $T_q \notin \mathbb{P}$ and $T_p \in \mathbb{P}$.
 $T_q \notin Clear(p)$  obviously holds  because of Step (1) in  Algorithm \ref{algorithm1}. 
 Considering two cases of $T_p$;  
 (a) In the case of $T_p \in Clear(p)$,  $T_q \to T_p$ does not hold  by $T_q \in \mathbb{P}$ in  the step (3) of  Algorithm \ref{algorithm1}; and  
(b) If $T_p \notin Clear(p)$, for $T_c \in Clear(p)$, then $T_c$ such that $T_q \to T_p \to T_c$ must exist by Step (3) of Algorithm \ref{algorithm1},  
however, $T_q \in Clear(p)$ must hold  by Lemma \ref{ssi4}, and  
thus  $T_p \notin Clear(p)$ reached a contradiction.  
Such $T_p$, $T_q$ cannot exist and contradict  the assumption because both cases of (a); and (b) were dismissed.    
Therefore, $\mathbb{P}$ is shown to be RSS and Algorithm \ref{algorithm1} is appropriate. 

\section{Implementations} \label{sec:implement}

% Figure 1 
\begin{figure*}[h]
 \begin{center}
  \includegraphics[bb=43 69 799 555,clip,width=0.85\linewidth]{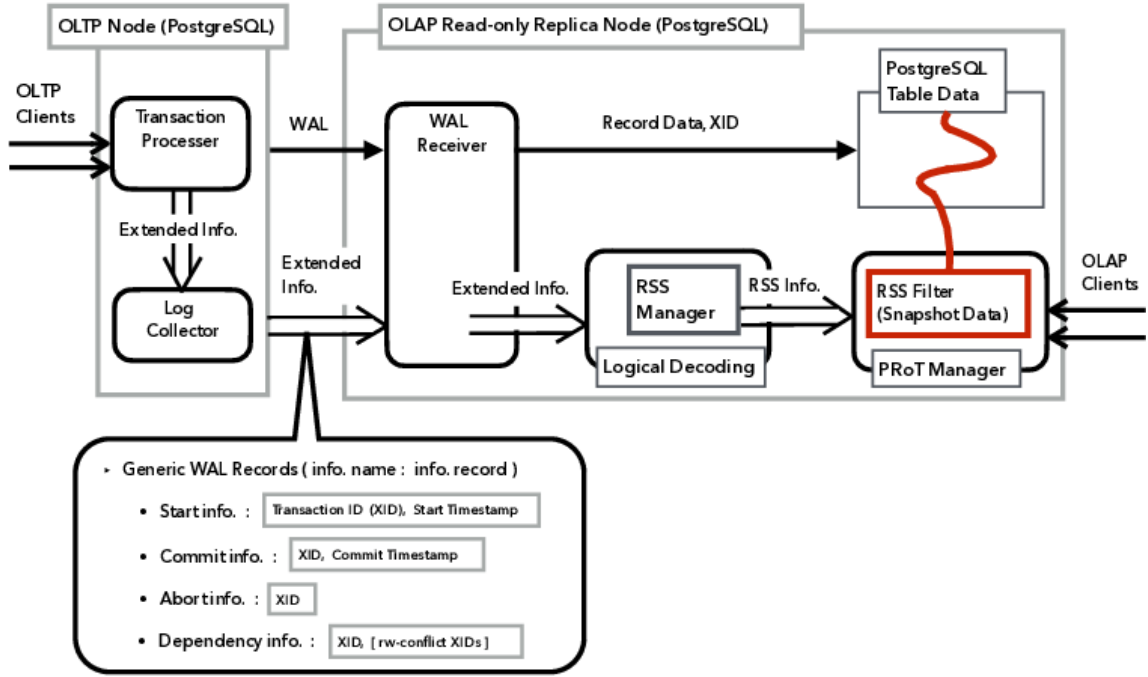}
  \caption{
  Multinode 
  %Decoupled OLTP/OLAP
   architecture overview. 
  Twin Arrows $\leftleftarrows$ represent clients.
  Arrow $\rightarrow$ represents existing processing flows of PostgreSQL.      
  Open arrow $\Rightarrow$ represents newly implemented processing flows.      
   }
  \label{fig:archi1}
 \end{center}
\end{figure*}

In this section, we introduce a prototype implementation that  practically uses Algorithm \ref{algorithm1}.  
 We implemented a prototype on top of PostgreSQL providing SSI, 
 because our algorithm assumes SSI  for serializability on OLTP. 
We implemented Algorithm \ref{algorithm1} on single and multinode architectures \cite{Raza2020} because  our model is applicable to both.
Although  unified  HTAP systems could have achieved OLAP's serializability derived from traditional transaction theories, 
 writer-/reader-abort or writer-/reader-wait  (rw-conflict blocking or Safe snapshots blocking)  have occurred    
  by read-only transaction participation. 
As we described in Section \ref{sec:2}, 
in the case of serializable isolation for a multinode HTAP architecture,  
the modern architecture must send the read-/write-set of concurrent transactions to each OLTP and OLAP system and synchronously check the read-/write-set,      
  or need to stop starting write transactions at fixed time intervals and let OLAP replica the wait the consistent view.  
In fact, 
 the idea of safe snapshots  for read-only replica is proposed in \cite{PortsDan2012}, but it has not been implemented yet.
PostgreSQL implementation uses the safe snapshots in the single-node architecture only, but 
 the method 
 cannot always complete within the fixed time interval and have to wait unexpectedly long times.

Our model  %improved from traditional MVCC theories 
  proves that the additional read-only transactions can participate in the existing set of transactions ensuring serializability:    
systems adopting the model do not forcibly abort or wait OLTP transactions and OLAP read-only transactions.     
We considered that our HTAP-compatible model could   refrain from deteriorating performance of multinode storage HTAP systems. 
Hence, we implemented the multinode architecture based on asynchronous log-shipping replication  PostgreSQL offers  and illustrated the implementation in Section \ref{impli}.

\subsection{ Multinode architecture} \label{impli}  
 %configuration 
 % for PostgreSQL} \label{impli} 

%
Our implementation is based on PostgreSQL version 12.0  as shown in Figure \ref{fig:archi1}.    
We list in the following sections the components or custom logic that play important roles to integrate  our model into PostgreSQL SSI. 
To realize SSI theory,  
PostgreSQL detects rw-conflicts between transactions while any concurrent transactions are alive. 
{\bf Dependency information.} 
Our implementation collects the detected outgoing rw-dependencies and writes the dependencies as a write-ahead log (WAL) record with self-transaction IDs to construct RSS  immediately after the reader transactions with outgoing edges are committed;   
the log collector shown in Figure \ref{fig:archi1} extracts the outgoing rw-dependencies that are relevant for our algorithm.
The collected dependencies about a transaction were written  by using logical messages that PostgreSQL offers  
as generic WAL records. 
Each of the direct conflicts was expressed as an array of writer transaction IDs in the generic WAL record. 
% written by the transaction. 

{\bf Start/End information.} 
In addition,  commit, start and abort information about when each transaction had started (induced by the first operation) and ended was appended to WAL.  
The written commit information was needed to construct the set $Done(p)$ of transaction schedules from a regular  
 % given arbitrary current prefix 
   WAL scan 
   on the OLAP side. 
   % through scanning WAL. 
%
The start/commit information was  required to create $Clear(p)$ on the prefix from Active transactions that   had started but not ended (committed or aborted) at the time when WAL inspection was executed.  
For the management of Active transactions, the abort information was used to exclude transactions from Active ones.

{\bf OLTP read-only transactions.}
Although  PostgreSQL implementation does not require a read-only transaction to have the transaction ID assigned, 
we assigned one for read transactions as well because   
the DBMS does not know beforehand if the transactions are read-only. 
As a result, the OLTP side was implemented  to collect dependencies and  start/end information about read-only transactions as well as write transactions.

{\bf Versions Preservation.} 
The logical messages we appended to WAL for constructing RSS were sent by an asynchronous commit option of streaming replication.
PostgreSQL is implemented to preserve multiversions  from overwriting previous versions of data items.
To maintain the old version, updated transactions create a new tuple adhered self-transaction ID and install the tuple in  storage. 
Stale tuples are deleted  % on read-only replica
 by two types of garbage collections called vacuum and heap-only-tuple (HOT). 
The vacuum is explicitly initiated by a user's call,  but  HOT is implicitly initiated despite the user's intention. 
Our implementation needed to maintain the stale tuples from HOT by sending Active transaction alive information from read-only replica called hot-standby-feedback in PostgreSQL configuration.  
%

% {\bf RSS construction invoker.} 
%
We use logical decoding that is offered by PostgreSQL  for reading the information in generic WAL records.  
We reformed the read replica on the streaming replication environment to be able to use the logical decoding.  
The logical decoding was executed from an  external processing module at fixed intervals. 
RSS construction invoker is the external module that  keeps snapshots and replaces them with RSS as transactions.  
Regularly transaction module executes UDFs that we created to decode the generic WAL, manage dependency graph,  and construct RSS.

{\bf RSS manager.} 
The decoded information was used to construct the brief transaction history on PostgreSQL shared memory. 
The history sequence consists of start and commit times of transactions that have been started or committed by the time the invoker transactions are executed. 
The RSS manager constructs Active, Done, Clear transactions using a hash table of transaction IDs by  scanning the history.  
Active transactions contain transactions having  start information only, while on  
 the previous prefix of the history, the previous Active transactions are transformed to Done  or  Clear  when commit  information  existed on the time scanning WAL.  
If abort info had arrived at the time of WAL inspection, 
 transaction IDs transfer from Active transactions to the set of transaction IDs for garbage collection. 
Clear and Done transactions are carefully managed for the RSS construction algorithm.
% described in Section \ref{SHTAP}.  
%

In addition, decoded dependency information is required to construct RSS, 
thus, we implemented a dependency graph on the shared memory of PostgreSQL to preserve the sent dependency information  as a vertex of the transaction ID and a path of the dependencies. 
If the dependency graph could discover paths from Clear transactions to Done transactions, 
 RSS  contains the discovered transaction IDs.     
Prepared as UDF, these RSS construction and dependency graph operations are called in a snapshot replacing transaction RSS construction invoker offers.

{\bf PRoT manager.} 
To replace snapshots with RSS, the snapshot-preserving transactions called from the RSS construction invoker must  keep executing by the time the next RSS construction ended. 
The RSS and dependency graph construction operations were implemented in PRoT manager as UDFs. 
The PRoT manager calls  RSS and dependency graph construction operations and receives the current RSS information as a snapshot data of PostgreSQL from the RSS manager. 
The RSS snapshot data are exported to transactions on the read-only replica. 
% 
%Such transactions could have read the versions created by the most recent committed transactions in RSS.  
%

\subsection{Single-node architecture} \label{setup_single}

Although our algorithm can be efficiently applied to the multinode architecture,   
comparative approaches achieving serializability have not existed on the multinode  architecture: 
HTAP systems are used to put off serializable isolation for better performance. 
In addition, 
the idea of safe snapshots on read replica in \cite{PortsDan2012} was not implemented in PostgreSQL.   
Because comparative methods like SSI and SSI + Safe Snapshots had been only applied in the single-node architecture,    
for evaluations of performance overheads from aborts and waits for serializability,  
we implemented our algorithm in single-node architecture. 
%

%Figure 3 
\begin{figure}[ht]
\centering
\includegraphics[bb=92 166 755 429,clip,width=1.0\linewidth]{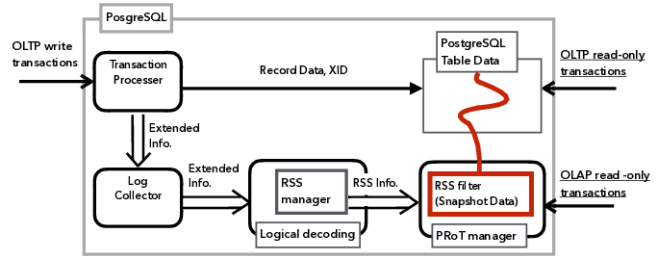} % 92 166 755 429
\caption{ Single-node  architecture overview  
%;  hybrid OLTP/OLAP workload analytical queries derived from TPC-H were executed on RSS. 
 } 
\label{fig:unifiedSys}
\end{figure}

The other system we implemented is summarized in Figure \ref{fig:unifiedSys}. 
%\noindent 
%
This single-node system was mostly based on the multi-node architecture in previous Section \ref{impli} without using log-shipping replication.
Versions that RSS contained had been firmly preserved by the snapshot-preserving transactions of RSS construction invoker until the next RSS constructed.  
This architecture is needed to recognize that transactions are read-only transactions for the use of RSS. 
Because analytical use queries like TPC-H benchmarks are known as read-only in advance,    
 we modified such benchmark queries by appending a PostgreSQL command that a transaction is specified as read-only. 
In addition, this implementation was modified to execute a transaction that could read RSS if a read-only flag in PostgreSQL  exists.   
Regarding OLTP read-only transactions as presented in the TPC-C benchmark, 
 RSS is not used but read-only anomalies do not arise by SSI as OLTP protocols,   
 because we considered that the OLTP side would be basically used as processing write transactions.

\section{Evaluation} \label{eval}

In this section,  
we evaluate the two types of prototype systems we proposed in Section \ref{sec:implement}.   
%
%{\bf Workload.}  
Our systems are evaluated under a CH-BenCHmark \cite{Cole2011} of OLTP-Benchmark \footnote{ \url{ https://github.com/oltpbenchmark/oltpbench} }
 \cite{Difallah2013OLTPBench}. 
OLTP-Benchmark implements CH-BenCHmark derived from TPC-C and TPC-H benchmarks  to 
 evaluate DBMSs designed to serve both OLTP and OLAP workloads.      
 All experiments were conducted on scale factor 100 (SF100; 100 warehouses on CH-BenCHmark) in the benchmark test.  
 Each test duration was 5 minutes and the warmup duration was 60 seconds after the initial data load of SF100 finished; 
the total run time was about 25 minutes.   
Comparison systems  were set up for each of evaluation purpose associated with our two systems as follows:    

\begin{itemize}
\item  
First, we investigated the abort rate and performance in comparison to both of PostgreSQL's SSI and safe snapshots that PostgreSQL can only support each serializable method in a single-node  architecture.  
We denote that applying safe snapshots (read-only deferrable transaction) in TPC-H queries is  "SSI+SafeSnapshots."   
% 
%Similarly, 
Executing OLTP transactions and OLAP queries under SSI  is denoted by "SSI."     
Our prototype system, where analytical queries derived from TPC-H were executed under RSS, is denoted by "SSI+RSS."   

\item 
Second, we test the performance overhead of RSS construction against SI that PostgreSQL offers as   repeatable read isolation levels in a read-only replica.   
Each primary and read-only replica was launched by PostgreSQL instances on two servers as  multinode architectures in this Section \ref{eval}.  
We denote that applying SI in the read-only replica is "SSI+SI." 
On the other hand, our system architecture, where the primary node uses SSI and the replica node applies RSS, is denoted by "SSI+RSS."  

\end{itemize}

\noindent 
{\bf Environment.}    
The experiments of single-node architectures were run on a client server publishing transactions of the CH-BenCHmark and a database server executing the transactions.  
The client server was equipped with four Intel(R) Xeon(R) Platinum 8176 CPUs with 2.10 GHz processors, having 28 physical cores  (32-KB L1i + 32-KB L1d cache, 1024-KB L2 cache, and 39.424-MB L3 shared cache) and 224 logical cores, 512 GB of DRAM and 440 GB of SSD.
The database server with two Intel(R) Xeon(R) Platinum 8176 CPUs clocked at 2.10 GHz with 28 physical cores (32-KB L1i + 32-KB L1d cache, 1024-KB L2 cache, and 39.424-MB L3 shared cache)  has 112 logical cores, 1 TB DRAM and 440 GB SSD.
In the case of the experiments of multinode architectures, PostgreSQL primary and replica  were  constructed on two database servers.  
All the experiments were run on these servers with the Ubuntu 18.04.3 OS and PostgreSQL 12.0. 
%

% Figure 5, 6
\begin{figure*}[h]
\begin{tabular}{c}
  %\begin{minipage}[b]{.100\linewidth}
  \begin{minipage}[t]{0.85\linewidth}
  \centering
    \includegraphics[bb=0 0 574 217, keepaspectratio,clip, width=0.85\linewidth]{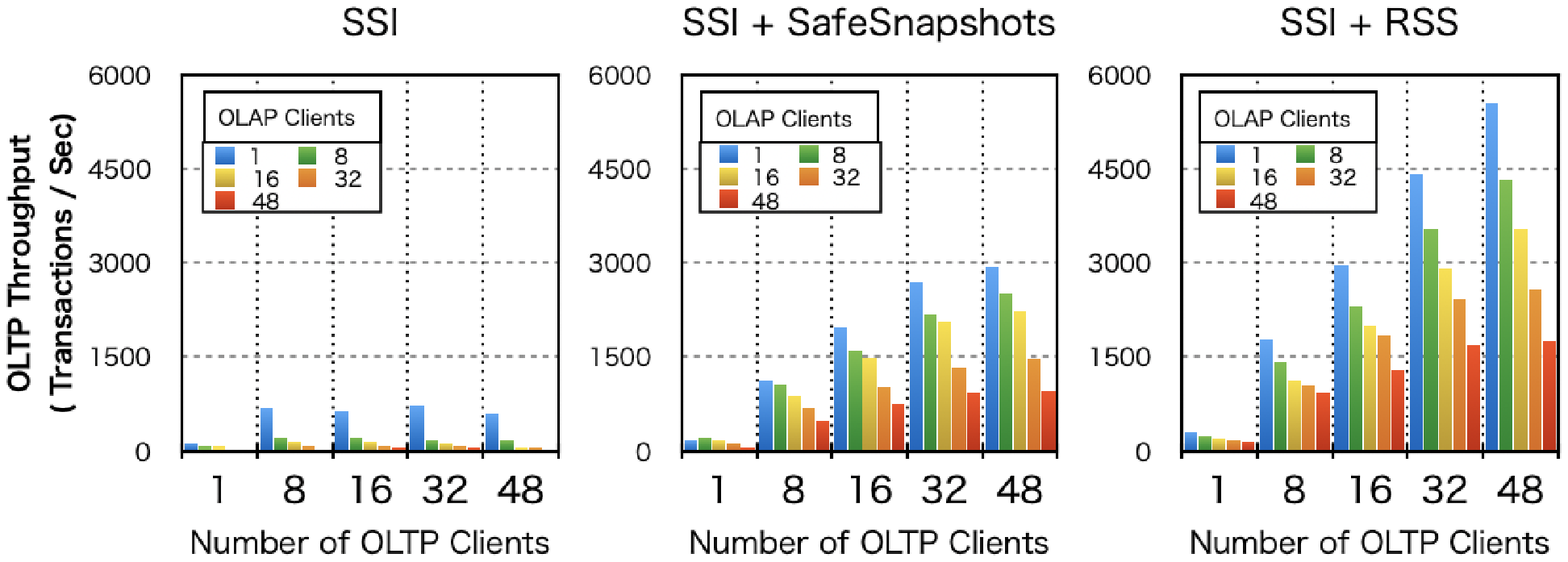}
    \vspace{-8pt}
    \caption{OLTP throughput  on single-node architecture}
    \label{OLTP_single}
 \vspace{10pt}
  \end{minipage} \\
  \begin{minipage}[t]{0.85\linewidth}
  \centering
    \includegraphics[bb=0 0 574 217,keepaspectratio,clip, width=0.85\linewidth]{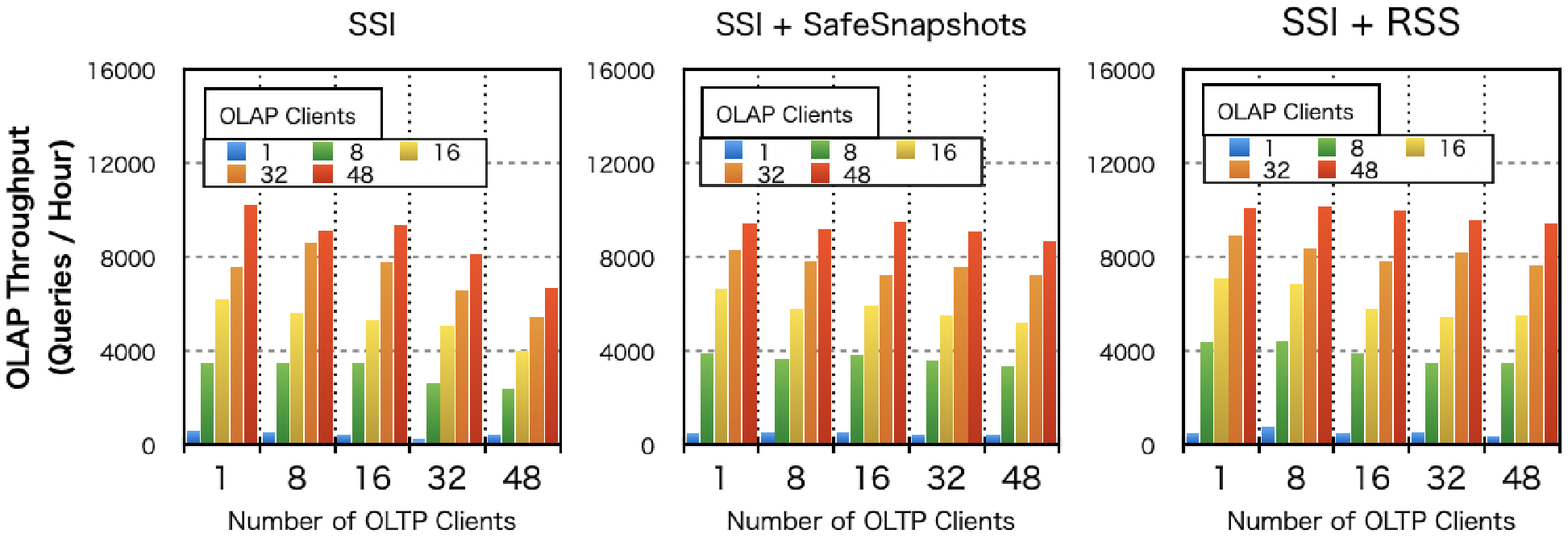}
    \vspace{-8pt}
    \caption{OLAP  throughput   on single-node architecture}
    \label{OLAP_single}
    \vspace{10pt}
  \end{minipage} \\
  \begin{minipage}[t]{0.85\linewidth}
  \centering
    \includegraphics[bb=0 0 574 217,keepaspectratio, clip, width=0.85\linewidth]{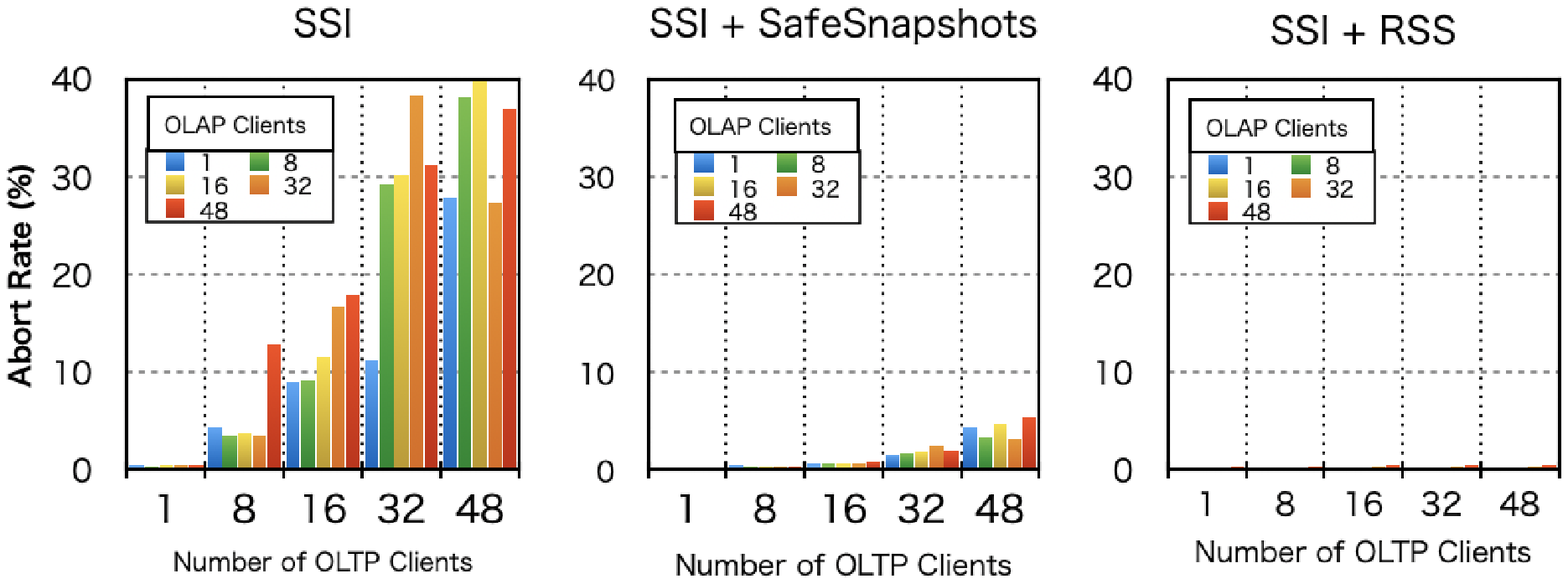}
    \vspace{-8pt}
    \caption{Abort rate  on single-node architecture}
    \label{Abort_single}
    \vspace{10pt}
  \end{minipage}
  \end{tabular}
  %\caption{Performance impact of each serializability method on unified architecture} 
\end{figure*}

\subsection{Serializability  impact on HTAP 
  }   \label{eval:single}
% Comparison with SSI and SSI+Safe Snapshots  at single-node}

%
We  experimentally compared with serializable methods of the SSI and SSI+SafeSnapshots under the hybrid OLTP and OLAP workloads.  
The serializability of SSI+RSS implementation of Section \ref{setup_single} was confirmed by reading the  previous version and using the example queries Ports and Grittner \cite{PortsDan2012} illustrated read-only anomaly in PostgreSQL.   
Reader-wait of safe snapshots (SSI+SafeSnapshots) serializability was also confirmed by the example of read-only anomalies: 
a read-only transaction waits to finish the concurrent transaction in the example.  
Such read-only deferrable transactions might affect concurrent write transactions in OLTP as much as OLAP.
Although these comparison methods might be affected by the validation cost of serializability,  
 we considered that SSI+RSS can save such costs derived from read-only transaction participation.  
Therefore, we investigated  
  the performance of OLTP and OLAP respectively, while varying the number of OLTP and OLAP clients. 
Figure \ref{OLTP_single} shows the transition of average OLTP performance executing CH-BenCHmark  
 when OLTP and OLAP clients increased to 48 from 1.  
  The x-axis shows the number of OLTP clients.  
The y-axis shows the OLTP transactions per second.  %(TPS).       
Similarly,  Figure \ref{OLAP_single} shows the transition of average OLAP performance when transactional clients and analytical clients were increased to 48 from 1. 
The y-axis shows the OLAP queries per hour. %(QPH). 
Figure \ref{Abort_single} shows the abort rate measured in the same experiments. 
In Figure \ref{Abort_single},  we aggregated transactions of aborts and retries of serialization failures that  resulted from the benchmark log  and summarized the rate as the quotient of both OLTP transactions and OLAP queries. 
%

% overall 
As shown in Figure \ref{OLTP_single},  
 %tendency as a whole of 
  the experiments show   
 % on the single-node server 
 %keeps 
 decreasing OLTP throughput on increasing the number of OLAP clients.  
On the other hand, OLAP throughput basically did not change in Figure \ref{OLAP_single} even though    the number of OLAP clients increased.  
Figure \ref{Abort_single} also  exhibits a similar tendency.  
%as a whole.  
%
%
% 
In particular, the SSI abort rate of 48 OLTP clients was  % caught
   about 35\% as abort transactions occurred even if one of the OLAP clients participated.   
Under CH-BenCHmark,    
  the OLAP read-only transactions were not aborted in exchange for aborting OLTP write transactions.  
Because dangerous structure tends to capture the long-running scan-heavy OLAP transactions as first reader transaction forming two successive rw-dependencies in contrast to the short-running OLTP write transactions.  
 OLTP throughput decreases by writer-aborts from read-only transaction participation. 
 In addition, 
  SSI is difficult for multi-node HTAP systems that cannot  efficiently execute such writer-abort described in Section \ref{sec:roa}.  
%

%({\bf Safe snapshots evaluation.}) 
%
 Consequently, SSI+SafeSnapshots applying read-only optimized SSI of PostgreSQL \cite{PortsDan2012}  could improve even better than SSI of OLTP and OLAP throughput. 
Because   
 read-only transactions could avoid such writer-aborts and rw-dependency checks in SSI validation  (taking the advantage of reducing the cost of SIRead Lock)  by reading snapshots ensuring beforehand serializable history,   
%
%Therefore, 
 the overall  abort rate under SSI+SafeSnapshots results was lower than SSI that of SSI.       
 In addition,  the OLAP throughput was less affected    
  in the face of 
 the reader-wait problem that read-only deferrable transactions must wait  
  %for executing queries
   by the time the safe snapshots are constructed.   
% 

%RSS
%
Figure \ref{OLAP_single} shows that the OLAP throughput of SSI+RSS was about 10\% higher on average  than SSI+SafeSnapshots on the runs from 1 to 48 OLAP clients. 
To realize efficiently serializable HTAP application,  
 RSS can  ensure serializability without reader-/writer-aborts by  read-only transaction participation in concepts that constructsa serializable snapshot  beforehand (wait-free snapshot read).   
However, the RSS OLAP throughput shows similar performance with SSI+SafeSnapshots.   
Compared with safe snapshots construction methods, the RSS construction method described in Section \ref{model} can execute  OLAP read-only transactions,  ensuring serializability  while executing concurrently long-running write transactions on the OLTP side.    
In the case of CH-BenCHmark having workload features of TPC-C like short-running write transactions, 
the write transactions started before the read-only deferrable transactions of SSI+SafeSnapshots can finish within short time.   
 Although 
 SSI+RSS collects rw-conflict and start/end information of extended information,  
 the snapshot construction overhead is considered to be lower than SSI+SafeSnapshots. 
As shown in Figure \ref{Abort_single}, the SSI+RSS overhead was lower than SSI+SafeSnapshots overhead in abort rate when OLTP clients increased. 
%the overheads were shown in abort rate when OLTP clients increased, SSI+RSS was lower than SSI+ SafeSnapshots.  
% 
SSI+RSS OLTP throughput was about 20\% higher than SSI+SafeSnapshots in the experiments.  
 of 48 OLTP clients. 
SSI+RSS  could improve OLTP performance despite the overheads collecting extended information for RSS construction.

\subsection{RSS handling cost for SI-based replica 
% installation cost for SI-based  replica construction     
% analysis 
  } 

\begin{figure}[h]
\begin{tabular}{l}
  \begin{minipage}[l]{1.0\linewidth}
 % \centering
    \includegraphics[clip,bb=0 0 510 228,keepaspectratio, width=1.0\linewidth]{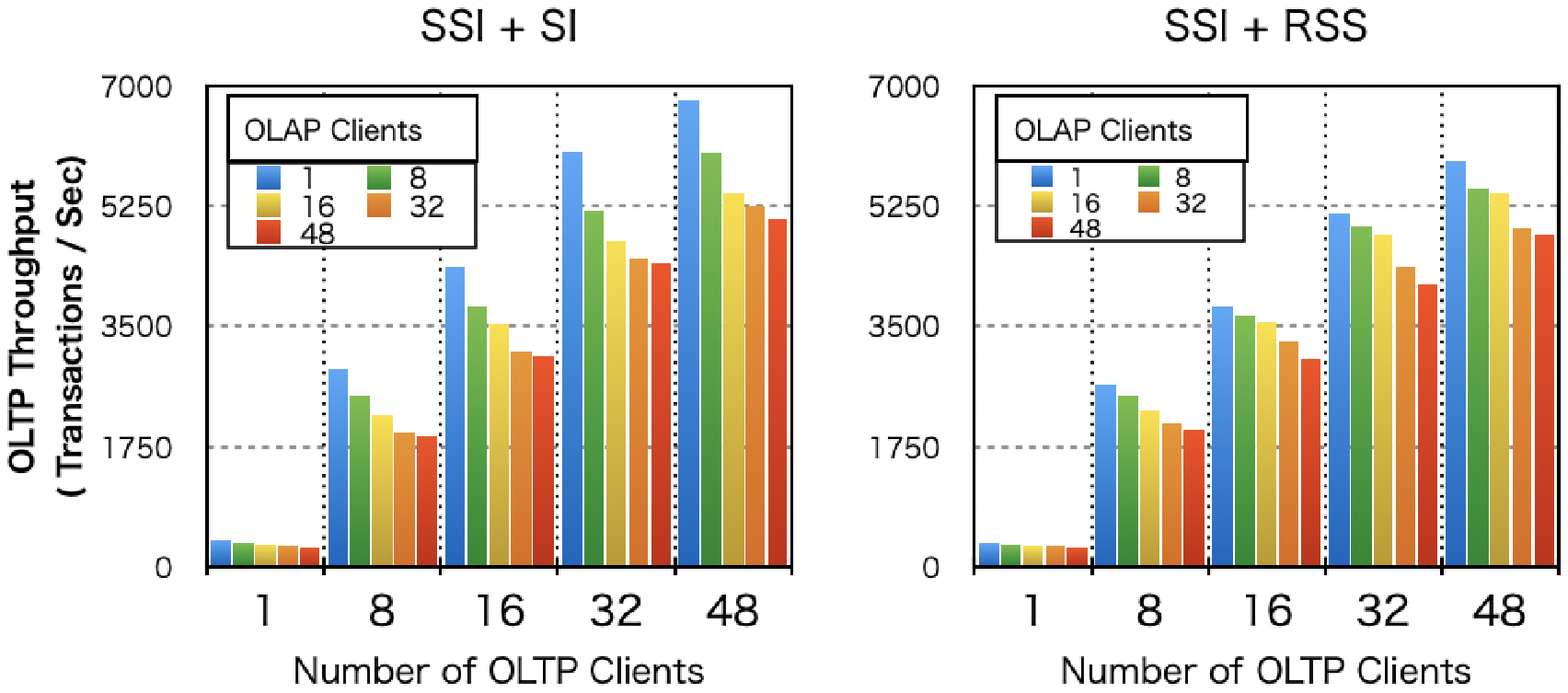}
    \vspace{-1pt}
    \caption{OLTP throughput of %OLTP engine
     on multinode  architecture  }
    \label{OLTP_multi}
 \vspace{10pt}
  \end{minipage} \\
  \begin{minipage}[l]{1.0\linewidth}
 % \centering
    \includegraphics[keepaspectratio,clip, bb=0 0 509 219, width=1.0\linewidth]{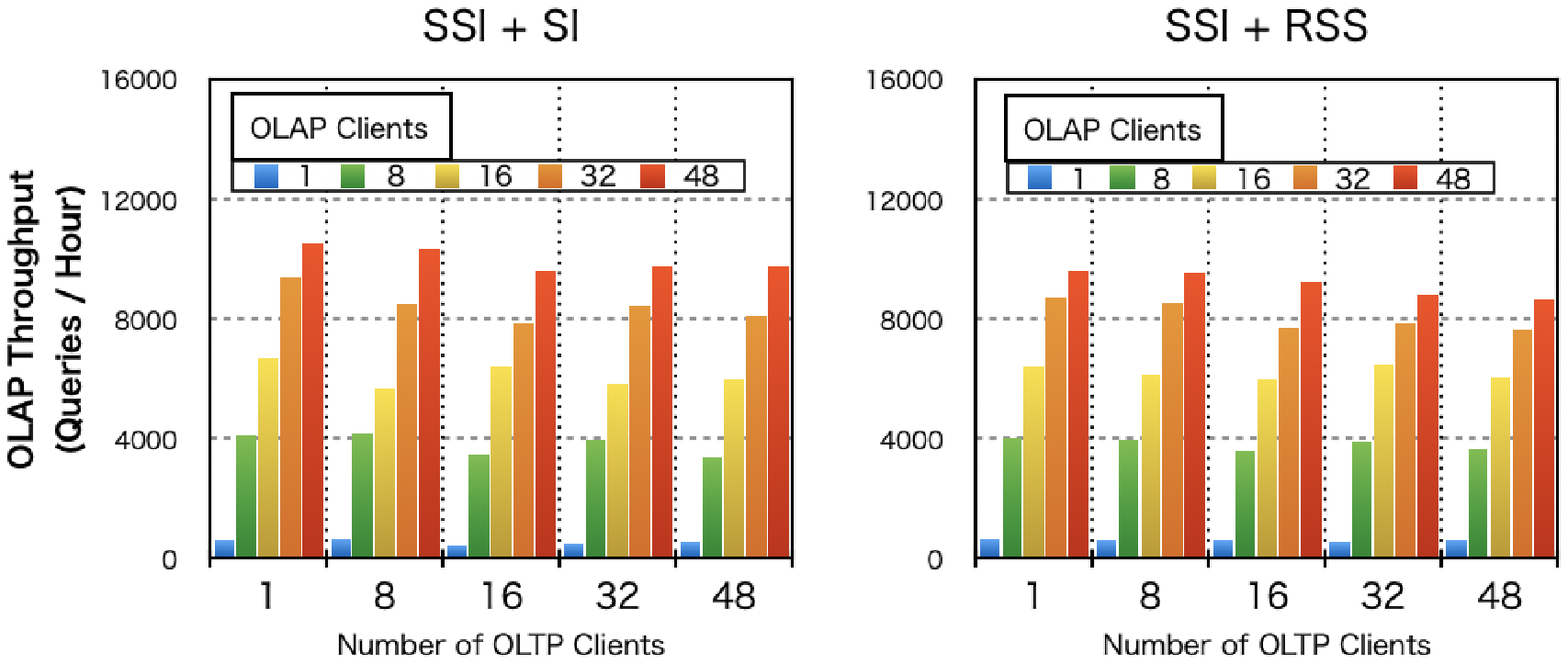}
    \vspace{-1pt}
    \caption{OLAP throughput  
    % of OLAP engine 
    on  multi-node  architecture  }
    \label{OLAP_multi}
    \vspace{10pt}
  \end{minipage} \\
  \begin{minipage}[l]{1.0\linewidth}
 % \centering
    \includegraphics[keepaspectratio, clip, bb=0 0 509 219,width=1.0\linewidth]{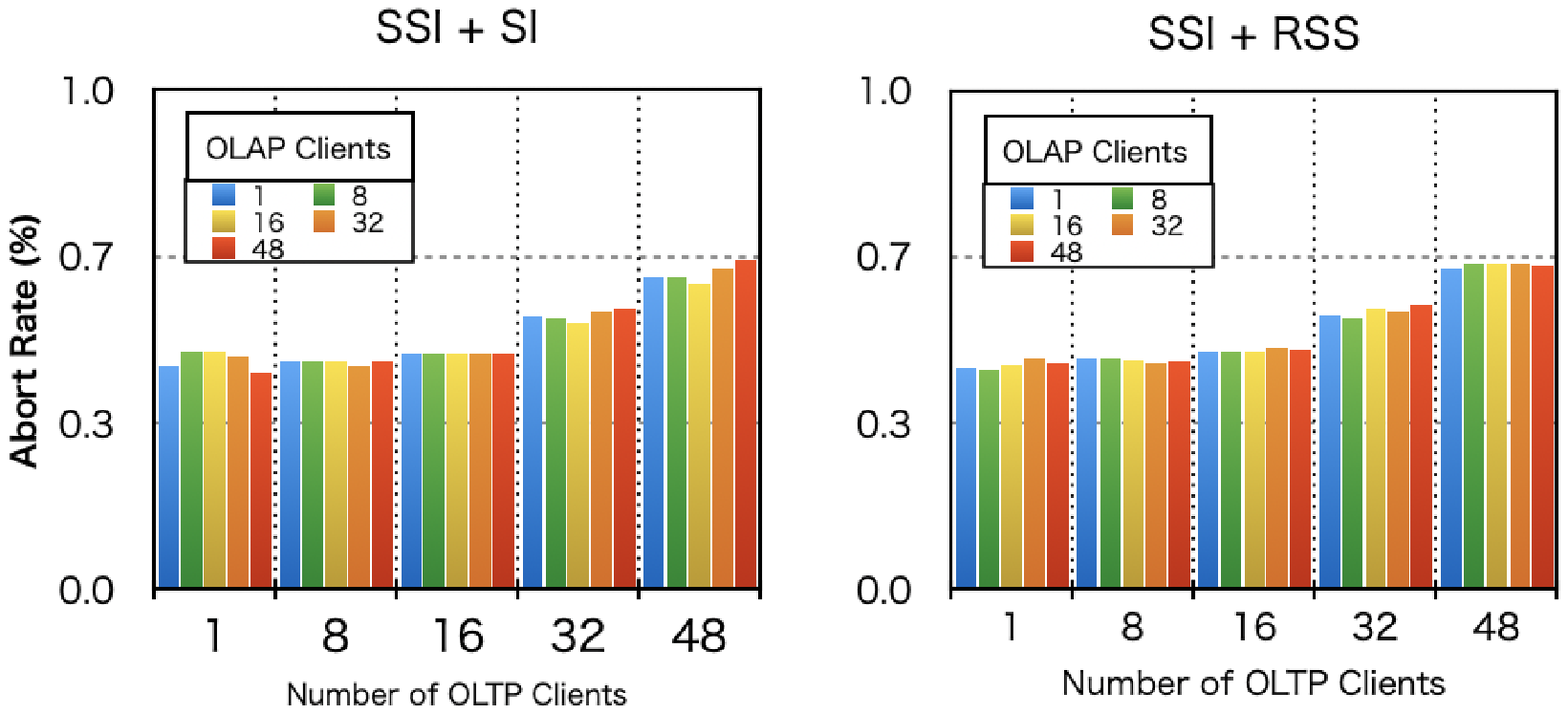}
    \vspace{-1pt}
    \caption{ Abort rate on  multinode  architecture }
    \label{Abort_multi}
    %\vspace{-1}
  \end{minipage}
  \end{tabular}
 % \caption{Performance impact of each serializability method on unified architecture} 
\end{figure}

In the following experiment, we tested RSS construction overheads in the case of a multinode architecture.   
In the same way as single-node experiments, we examined the OLTP and OLAP throughput of CH-BenCHmark. 
%
%However, multinode architecture systems could have not achieve serializability of OLAP, 
We evaluated our system that could achieve serializability in comparison with SSI+SI, which could have not achieve serializability of OLAP replica by read-only anomalies, for      
% 
%The evaluation we describe in this section
 investigation of our serializability assurance overhead.    
In the same way as single-node experiments, 
 Figure \ref{OLTP_multi} shows the OLTP throughput of SSI+SI and SSI+RSS,  
 Figure \ref{OLAP_multi} shows the OLAP throughput, 
 and  Figure \ref{Abort_multi} shows the abort rate. 
%

%
% Figure 8,9,10

As in the single-node case in Section \ref{eval:single}, 
 on the results of SSI+RSS in Figure \ref{OLTP_multi}, show that  even if  the number of OLAP clients increases, OLTP performance  mostly does not fall.  
On the other hand,  the SSI+SI read-only replica executing OLAP queries under SI  does not need  serialization cost.    
Nevertheless,   
 OLAP performance of SSI+RSS is almost the same performance of SSI+SI because of wait-free snapshot read by constructing RSS  beforehand. 
On the OLTP side,  SSI+SI is always better than the OLTP throughput of SSI+RSS.
We set hot\_standby\_feedback and replication slot PostgreSQL offers the version preservation of replica  for correct evaluation of SSI+SI  because OLAP query results in errors that  try to read deleted versions,  and we make the same settings of the replica equal to SSI+RSS.   %equal to
We consider  OLTP degradations of SSI+SI and SSI+RSS to be caused by preserving old versions,  disabling HOT mechanism utilization on PostgreSQL.
In particular, by reading the versioned tuples in the direction from the oldest to the newest in  the transaction processing of  PostgreSQL, OLAP clients would result in OLTP performance deterioration. 
Furthermore, the OLTP side of SSI+RSS forces each write transaction to  log extended information, which  caused performance deterioration.    
These factors show that the serialization cost of OLAP read-only transactions would be bigger than nonserializable analysis of SI in real time, because 
%, 
 SSI+RSS OLTP average throughput is 10 \%  lower   
  %1\% $\backsim$ \%15 
    than SSI+SI when experiments run on 48 OLTP clients. %s in Figure \ref{OLTP_multi}.
 In contrast, Figure \ref{OLAP_multi} and Figure\ref{Abort_multi}  show that the OLAP throughput and abort rate are almost the same in overall runs: 
 these results are considered that RSS achieves wait-/abort-free read with serializability by read-only transaction participation in real time.

\section{Conclusion} \label{conclusion} 

% Background  
% 
This paper addresses a missing part of  MVCC aspects of read-only transactions and version selection. 
In particular, we considered that serializability, where read-only transactions frequently participated, will be needed for database applications like HTAP.  
%
% Proposal 
%
We introduced the concepts of RSS and the construction algorithm to  achieve global serializability   meaning that OLAP read-only transactions can step in OLTP transaction histories by choosing previous versions.     

Our prototype systems based on PostgreSQL could apply the algorithm for HTAP architectures categorized into two types called  unified (single-node) and decoupled (multinode) systems. 
We evaluated the performance of our systems under OLTP and OLAP workloads. 
In the single-node system where serializable methods (SSI or SSI+SafeSnapshots) were offered,  
 we improved OLTP throughput by up to 20\%.  
In addition, the RSS construction overhead on a multinode system was restrained by 10\% in comparison with SSI+SI that causes anomalies on OLAP replica;   
 SSI+SI  cannot achieve serializability. 
 % and  as CC of a general multinode system, 
 %  by aborting dangerous structures or creating safe snapshots. 
  %because of  both OLTP and OLAP performance degradation.  
% by detecting dangerous structures and creating safe snapshots:  
%   
% the SSI+SI OLAP replica contains many dangerous structure  read only anomalies.  
% 
 In contrast, SSI+RSS achieve serializability and 
 the OLTP/OLAP throughput did not degrade much against the comparison methods on single-/multinode architecture, and the OLTP/OLAP abort rate also did not increase: 
these show that our approach can read snapshots wait-/abort-free even if read-only transactions were continuously executed. 
%
%In this paper, workloads including both OLTP and OLAP within a single transaction are not covered.
% 
%Our future work will aim to handle such workloads. 
% 
We will aim to handle workloads including both OLTP and OLAP within a single transaction in future work. 
Besides, we are going to study scale-out architecture and tiering capabilities of multinode servers for the RSS-based  performance enhancement because RSS can reasonably add serializable read-only replica nodes without consensus or synchronous commit algorithms.   

\section{acknoledgements}

%\begin{acks}
%
This work is based on results obtained from project JPNP16007, commissioned by the New Energy and Industrial Technology Development Organization (NEDO).  
%
% This work was supported by the [...] Research Fund of [...] (Number [...]). Additional funding was provided by [...] and [...]. We also thank [...] for contributing [...].
%\end{acks}

%\clearpage


\balance 
\end{document}